\documentclass[extra, mreferee]{gji}

\usepackage{epsfig}
\usepackage{amssymb, amsmath}
\usepackage{setspace}

\title[Marchenko for evanescent waves]
 { The Marchenko method for evanescent waves}

\author[Wapenaar]
{\small Kees Wapenaar\\
  Department of Geoscience and Engineering, Delft University of Technology, \\P.O. Box 5048, 2600 GA Delft, The Netherlands}
\pagerange{1--5}
\volume{999}
\pubyear{2020}

\begin{document}
\begin{spacing}{2}
\label{firstpage}

\maketitle

\begin{summary}
{\small With the Marchenko method, Green's functions in the subsurface 
can be retrieved from seismic reflection data at the surface.
State-of-the-art Marchenko methods work well for propagating waves but break down for evanescent waves. 
This paper discusses a first step towards extending the Marchenko method for evanescent waves and analyses its possibilities and limitations.
In theory both the downward and upward decaying components can be retrieved. The retrieval of the upward decaying component appears to be very sensitive to model errors, but 
 the downward decaying component, including multiple reflections, can be retrieved in a reasonably stable and accurate way.
The reported research opens the way to develop new Marchenko methods that can handle refracted waves in wide-angle reflection data.\\
\mbox{}\\}
\end{summary}


\section{Introduction}

Building on the single-sided autofocusing method of \cite{Rose2002IP}, \cite{Broggini2012EJP}  proposed a data-driven method to retrieve the Green's function
inside a layered medium from the seismic reflection response at the surface. This method, which is based on the Marchenko equation, has been extended for laterally varying media
and used for imaging the subsurface without artefacts related to internal multiple reflections \citep{Wapenaar2014GEO, Ravasi2016GJI, Staring2018GEO}.
Current Marchenko methods only handle propagating waves, which for most practical applications is acceptable. 
However, in reflection experiments with large horizontal offsets, which may include refracted arrivals, evanescent waves play a significant role. 
This paper discusses a first step towards extending the Marchenko method for evanescent waves and analyses its possibilities and limitations.

\section{Propagation invariants}

We review propagation invariants for a horizontally layered lossless acoustic medium, which will be used for the derivation of representations for the Marchenko method in the next section.
The propagation velocity $c(z)$ and mass density $\rho(z)$ are piecewise continuous functions of the depth coordinate $z$.
In this medium, we consider a 2D space- and time-dependent acoustic wave field, characterized by $p(x,z,t)$ and $v_z(x,z,t)$, 
where $p$ is the acoustic pressure, $v_z$ the vertical component of the particle velocity, $x$ the horizontal coordinate and $t$ the time.
We define the temporal and spatial Fourier transform of $p(x,z,t)$ as
\begin{eqnarray}\label{eqA11}
\tilde p(s_x,z,\omega)=\int_{-\infty}^\infty\int_{-\infty}^\infty p(x,z,t)\exp\{i\omega (t-s_x x)\}{\rm d}t{\rm d}x,
\end{eqnarray}
where  $i$ is the imaginary unit, $\omega$ the angular frequency and $s_x$ the horizontal slowness. A similar definition holds for $\tilde v_z(s_x,z,\omega)$.
Throughout this paper $\omega$ is taken positive or zero. 
Since we use slowness $s_x$ (instead of wavenumber $k_x=\omega s_x$) as the spatial Fourier variable in $\tilde p(s_x,z,\omega)$, the inverse temporal Fourier transform is defined per $s_x$-value as
\begin{eqnarray}\label{eqA11invb}
&&p(s_x,z,\tau)=\frac{1}{\pi}\Re\int_{0}^\infty \tilde p(s_x,z,\omega)\exp(-i\omega \tau){\rm d}\omega.
\end{eqnarray}
Here $\Re$ denotes the real part and $\tau$ is the so-called intercept time \citep{Stoffa89Book}.
For $\tilde p(s_x,z,\omega)$ as well as $p(s_x,z,\tau)$, the wave field is propagating when $|s_x|\le 1/c(z)$ and evanescent when $|s_x|> 1/c(z)$.
For propagating waves, the local propagation angle $\alpha(z)$ follows from $s_x=\sin\alpha(z)/c(z)$. 
Everything that follows also holds for 3D cylindrically symmetric wave fields when the spatial Fourier transform is 
replaced by a Hankel transform and the horizontal slowness $s_x$ by the radial slowness $s_r$.

We consider two independent acoustic states, indicated by subscripts $A$ and $B$. The following combinations of wave fields in states $A$ and $B$, 
\begin{eqnarray}
&&\tilde p_A\tilde v_{z,B}-\tilde v_{z,A}\tilde p_B\label{prop1}
\end{eqnarray}
and
\begin{eqnarray}
&&\tilde p_A^*\tilde v_{z,B}+\tilde v_{z,A}^*\tilde p_B\label{prop2}
\end{eqnarray}
(with the asterisk denoting complex conjugation), are propagation invariants. This means that for fixed $s_x$ and $\omega$
these quantities are independent of the depth coordinate $z$ in any source-free region \citep{Kennett78GJRAS}.
A special case is obtained when we take states $A$ and $B$ identical: dropping the subscripts $A$ and $B$ in equation (\ref{prop2}) and multiplying by a factor $1/4$,  this yields
the power-flux density in the $z$-direction, i.e.,
\begin{eqnarray}
&&j=
 \frac{1}{4}\{\tilde p^*\tilde v_z + \tilde v_z^*\tilde p\}.\label{eqflux}
\end{eqnarray}
Next, we introduce pressure-normalized downgoing and upgoing fields $\tilde p^+$ and $\tilde p^-$, respectively, and relate these to the total fields $\tilde p$ and $\tilde v_z$, via
\begin{eqnarray}
&&\tilde p=\tilde p^+ + \tilde p^-,\label{eqcomp1}\\
&&\tilde v_z=\frac{s_z}{\rho}(\tilde p^+ - \tilde p^-).\label{eqcomp2}
\end{eqnarray}
Here $s_z(z)$ is the vertical slowness. For propagating waves it is positive real-valued or zero, according to
 \begin{eqnarray}\label{eqqreal}
&&s_z= +\sqrt{1/c^2-s_x^2},\quad\mbox{for}\quad s_x^2 \le 1/c^2(z),
\end{eqnarray}
whereas for evanescent waves it is positive imaginary-valued, i.e.,
\begin{eqnarray}\label{eqqimag}
&&s_z= +i\sqrt{s_x^2-1/c^2},\quad\mbox{for}\quad s_x^2 > 1/c^2(z).
\end{eqnarray}
For evanescent waves, $\tilde p^+$ and $\tilde p^-$ are downward and upward decaying (i.e., decaying in the $+z$ and $-z$ direction), respectively.
Substitution of equations (\ref{eqcomp1}) and (\ref{eqcomp2}) into equations (\ref{prop1})  and (\ref{prop2}) yields two additional propagations invariants  \citep{Ursin83Geo, Wapenaar89Geo}
\begin{eqnarray}
&&-\frac{2s_z}{\rho}\bigl(\tilde p_A^+\tilde p_B^- - \tilde p_A^-\tilde p_B^+\bigr)\label{prop3}
\end{eqnarray}
and 
\begin{eqnarray}
&&\frac{2\Re(s_z)}{\rho}\bigl((\tilde p_A^+)^*\tilde p_B^+ - (\tilde p_A^-)^*\tilde p_B^-\bigr)-\frac{2i\Im(s_z)}{\rho}\bigl((\tilde p_A^+)^*\tilde p_B^- - (\tilde p_A^-)^*\tilde p_B^+\bigr),\label{prop4}
\end{eqnarray}
respectively, where $\Im$ denotes the imaginary part. 
The second propagation invariant consists of two terms, of which only the first term is non-zero for propagating waves, whereas for evanescent waves only the second term is non-zero.
This second term was neglected in previous derivations of the Marchenko method.
In a layered medium, where tunnelling of evanescent waves occurs in thin high-velocity layers, the propagation invariant of equation 
(\ref{prop4}) switches back and forth between the first and the second term,
but its  value is the same in each layer.
Finally, for the special case that states $A$ and $B$ are identical we obtain for the power-flux density
\begin{eqnarray}
&&j=\frac{\Re(s_z)}{2\rho}\bigl(|\tilde p^+|^2-|\tilde p^-|^2\bigr)+\frac{\Im(s_z)}{\rho}\Im\bigl((\tilde p^+)^*\tilde p^-\bigr).
\end{eqnarray}
The first term quantifies the power-flux density of  propagating waves 
and the second term that of tunnelling evanescent waves in high-velocity layers.

\section{Representations for the Marchenko method}\label{sec3}

We use the propagation invariants of equations (\ref{prop3}) and (\ref{prop4}) to derive representations for the Marchenko method, analogous to \cite{Slob2014GEO} and \cite{Wapenaar2014GEO}, 
but extended for evanescent waves.
We consider a layered  source-free lossless medium for $z\ge z_0$. For state $B$ we consider a Green's function $\tilde G=\tilde G^++\tilde G^-$, with its source
(scaled with $-i\omega\rho$) just above $z_0$.
At $z_0$, the downgoing Green's function $\tilde G^+$ equals  $\rho(z_0)/2s_z (s_x,z_0)$ \citep{Aki80Book, Fokkema93Book}. The
wave fields $\tilde p_B^+$ and $\tilde p_B^-$ at $z_0$ (just below the source) and at $z_F$ (an arbitrarily chosen focal depth inside the medium) 
are given in Table 1. Note that $\tilde R^\cup (s_x,z_0,\omega)$ denotes the reflection response ``from above'' of the layered medium. 
For state $A$ we introduce a focusing function $\tilde f_1=\tilde f_1^++\tilde f_1^-$ in a truncated medium, which is identical to the actual medium above the focal depth $z_F$ 
and homogeneous below it. The downgoing focusing function $\tilde f_1^+(s_x,z,z_F,\omega)$ is defined such that, when emitted from $z=z_0$ into the medium, it focuses at 
$z_F$.  Its propagation to the focal depth $z_F$ is described by
$\tilde T^+(s_x,z_F,z_0,\omega)\tilde f_1^+(s_x,z_0,z_F,\omega)=\tilde f_1^+(s_x,z_F,z_F,\omega)$, where $\tilde T^+(s_x,z_F,z_0,\omega)$ 
is the downgoing  transmission response of the truncated medium
and $\tilde f_1^+(s_x,z_F,z_F,\omega)$ is the focused field at $z_F$. We could define $\tilde f_1^+(s_x,z_F,z_F,\omega)=1$, where $1$ is the Fourier transform of a temporal delta function.
However, in analogy with the downgoing Green's function at $z_0$, 
we define $\tilde f_1^+(s_x,z_F,z_F,\omega)=\rho(z_F)/2s_z (s_x,z_F)$, see Table 1. We thus obtain
\begin{eqnarray}
&&\tilde f_1^+(s_x,z_0,z_F,\omega)
=\frac{\rho(z_F)}{2s_z(s_x,z_F)}\frac{1}{\tilde T^+(s_x,z_F,z_0,\omega)}.\label{eqTinv}
\end{eqnarray}
Hence, the downgoing focusing function $\tilde f_1^+(s_x,z_0,z_F,\omega)$ is defined as a scaled inverse of the transmission response of the truncated medium.
The upgoing focusing function  $\tilde f_1^-(s_x,z_0,z_F,\omega)$ is the reflection response of the truncated medium to $\tilde f_1^+(s_x,z_0,z_F,\omega)$.
Since the half-space below the truncated medium is homogeneous, we have $\tilde f_1^-(s_x,z_F,z_F,\omega)=0$.

\begin{center}
{\noindent \it Table 1: Quantities to derive representations (\ref{eq145}) $-$ (\ref{eq148}).}
\begin{tabular}{||l|c|c|c|c||}
\hline\hline
& $\tilde p_A^+ (s_x,z,\omega) $ & $\tilde p_A^-(s_x,z,\omega) $ & $\tilde p_B^+ (s_x,z,\omega) $ & $\tilde p_B^-(s_x,z,\omega) $  \\
\hline
$z=z_0$  & $\tilde f_1^+(s_x,z_0,z_F,\omega)$ &$\tilde f_1^-(s_x,z_0,z_F,\omega)$ &  
$\frac{\rho(z_0)}{2s_z (s_x,z_0)}$ & $\frac{\rho(z_0)\tilde R^\cup (s_x,z_0,\omega)} {2s_z (s_x,z_0)} $\\
\hline
$z=z_F$  & $\frac{\rho(z_F)}{2s_z (s_x,z_F)}$ &$0$ & $\tilde G^+ (s_x,z_F,z_0,\omega) $ & $\tilde G^-(s_x,z_F,z_0,\omega) $ \\
\hline\hline
\end{tabular}
\end{center}

\mbox{}\\

The propagation invariants are now used to relate the quantities in Table 1 at $z_0$ to those at $z_F$.
From propagation invariant (\ref{prop3}) we obtain (for propagating and evanescent waves)
\begin{eqnarray}
&&\tilde G^-(s_x,z_F,z_0,\omega) + \tilde f_1^-(s_x,z_0,z_F,\omega)=\tilde R^\cup (s_x,z_0,\omega)\tilde f_1^+(s_x,z_0,z_F,\omega),\label{eq145g}
\end{eqnarray}
or, using the inverse Fourier transform defined in equation (\ref{eqA11invb}),
\begin{eqnarray}
&&G^-(s_x,z_F,z_0,\tau) + f_1^-(s_x,z_0,z_F,\tau)=\int_{-\infty}^\tau R^\cup (s_x,z_0,\tau-\tau')f_1^+(s_x,z_0,z_F,\tau'){\rm d}\tau'.\label{eq145}
\end{eqnarray}

Next we use propagation invariant (\ref{prop4}). First we consider propagating waves at $z_0$ and $z_F$. For this situation we only use the first term of this propagation invariant.
Substituting the quantities of Table 1 and applying the inverse Fourier transform of equation (\ref{eqA11invb}), we obtain
\begin{eqnarray}
G^+(s_x,z_F,z_0,\tau) - f_1^+(s_x,z_0,z_F,-\tau)=-\int_{-\infty}^\tau R^\cup (s_x,z_0,\tau-\tau')f_1^-(s_x,z_0,z_F,-\tau'){\rm d}\tau'.\label{eq146}
\end{eqnarray}
Next, we consider propagating waves at $z_0$ and evanescent waves at $z_F$. Equating the first term of propagation invariant (\ref{prop4}) at $z_0$ to the second term
at $z_F$, we obtain for the quantities of Table 1 (after an inverse Fourier transform)
\begin{eqnarray}
G^-(s_x,z_F,z_0,\tau) - f_1^+(s_x,z_0,z_F,-\tau)=-\int_{-\infty}^\tau R^\cup (s_x,z_0,\tau-\tau')f_1^-(s_x,z_0,z_F,-\tau'){\rm d}\tau'.\label{eq148}
\end{eqnarray}
Equations (\ref{eq145}) and (\ref{eq146}) were already known but equation (\ref{eq148}) is new.
It expresses the upward decaying part of the Green's function at $z_F$ in terms of the reflection response at the surface and focusing functions.
Note that two more relations can be derived for  evanescent fields at $z_0$, but these will not be discussed here.

\begin{figure}
\vspace{-.4cm}
\centerline{\epsfxsize=11cm \epsfbox{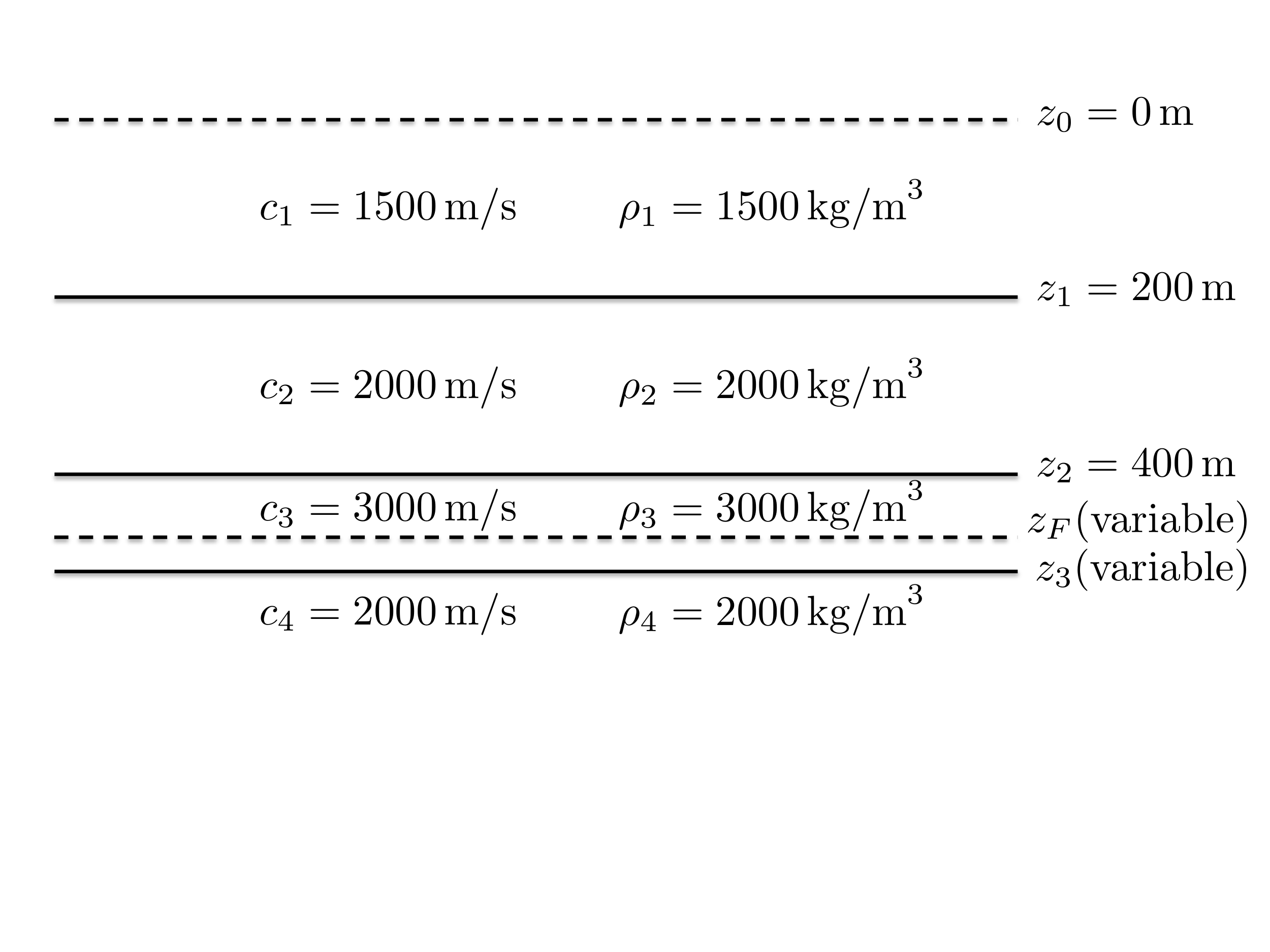}}
\vspace{-1.8cm}
\caption{Horizontally layered lossless acoustic medium.}\label{Fig1}
\end{figure}

\begin{figure}
\centerline{\epsfxsize=9 cm \epsfbox{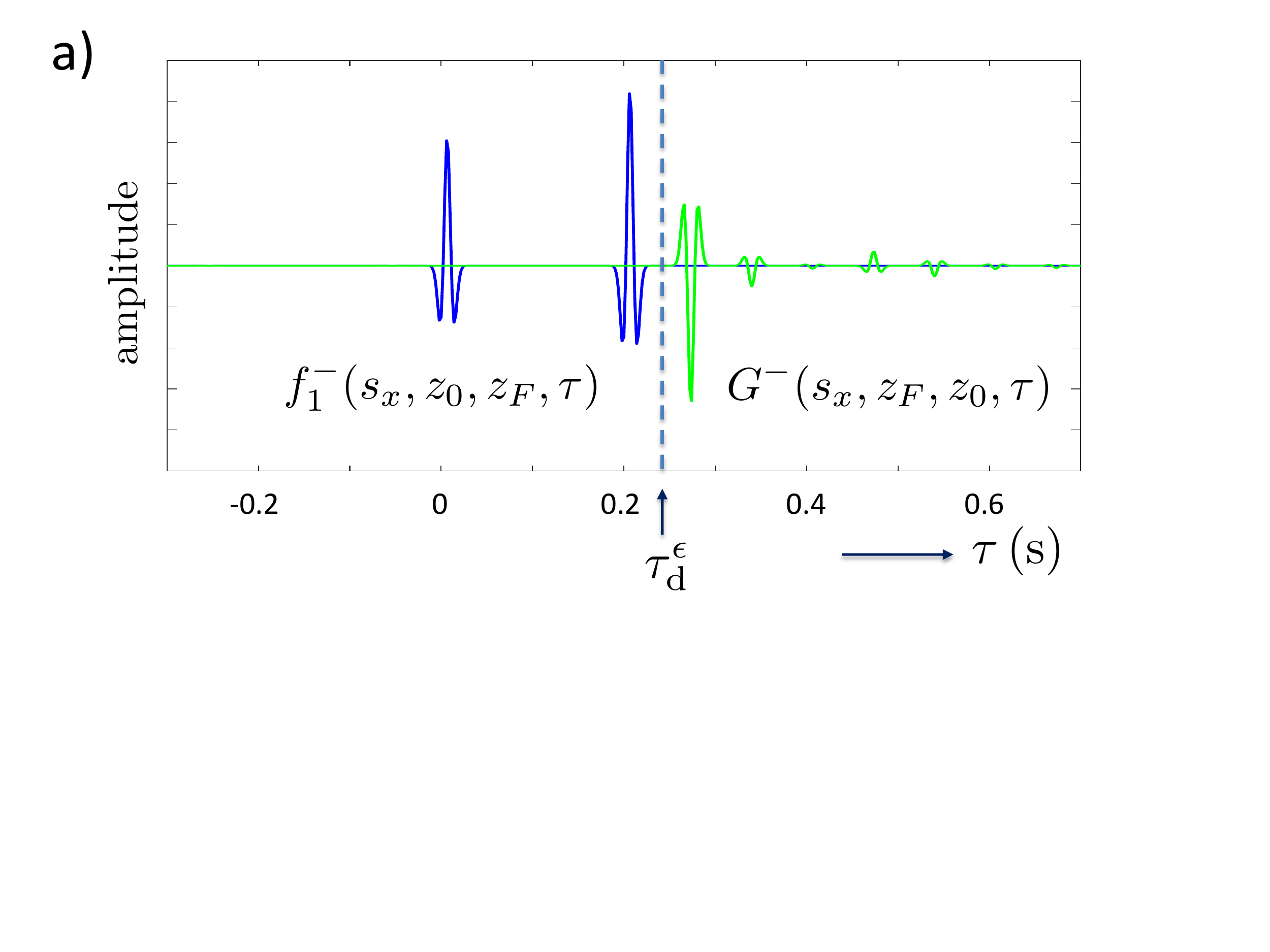}}
\vspace{-2.cm}
\centerline{\epsfxsize=9 cm \epsfbox{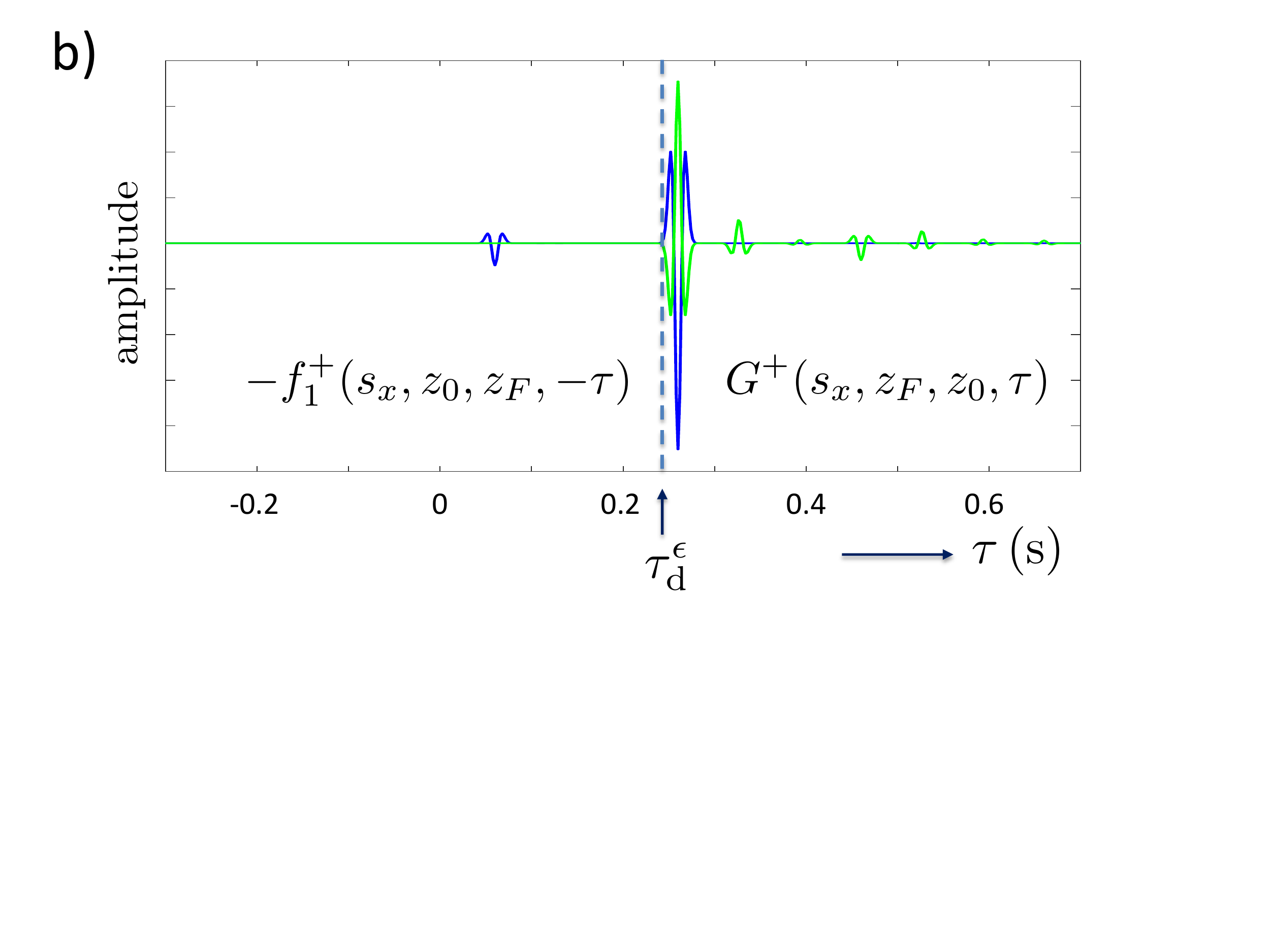}}
\vspace{-2.3cm}
\caption{Functions in the left-hand sides of (a) equation (\ref{eq145}) and (b) equation (\ref{eq146}), for propagating waves at $z_0$ and $z_F$. 
}\label{Fig2}
\end{figure}

\begin{figure}
\centerline{\epsfxsize=9 cm \epsfbox{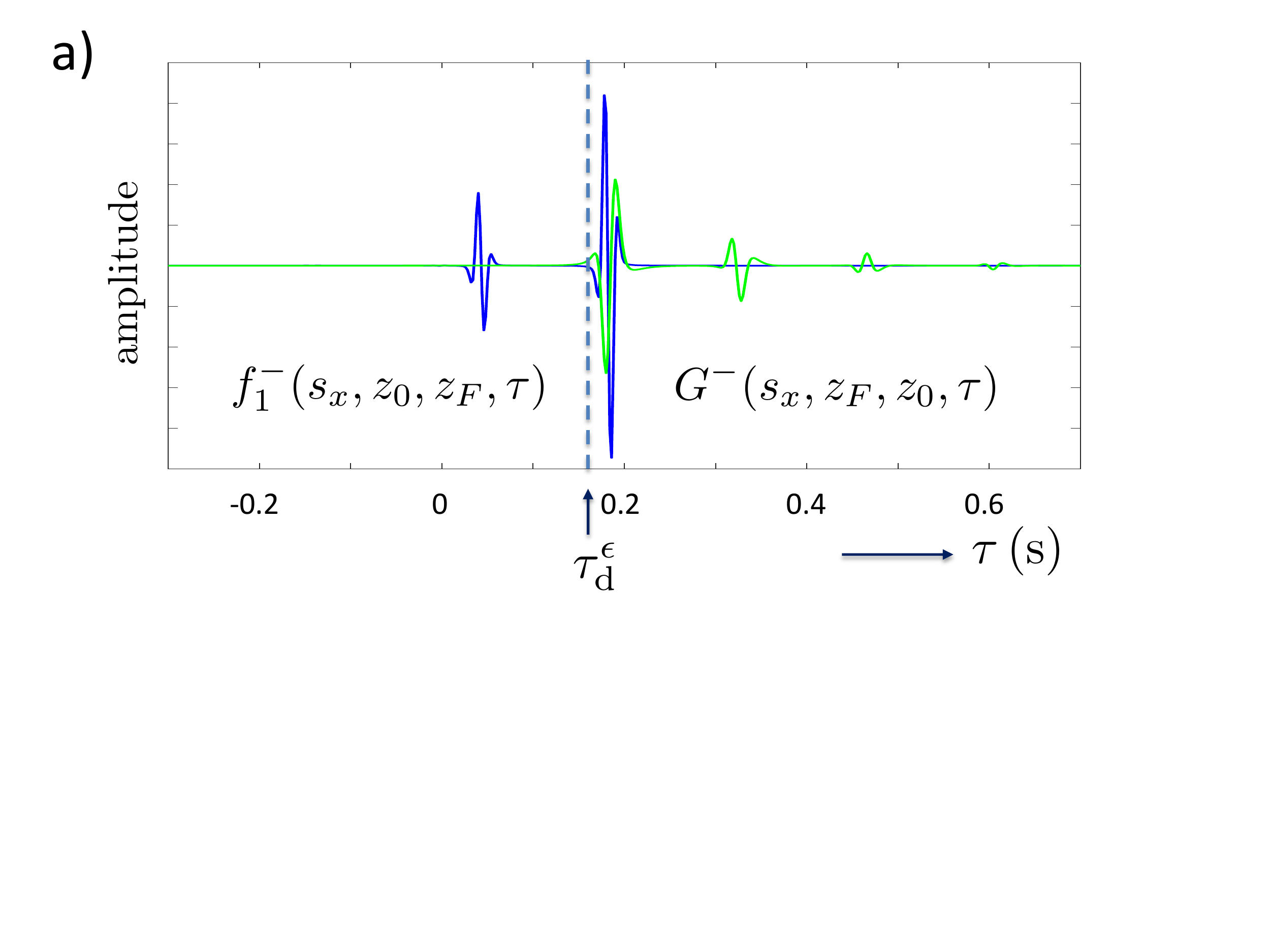}}
\vspace{-2.cm}
\centerline{\epsfxsize=9 cm \epsfbox{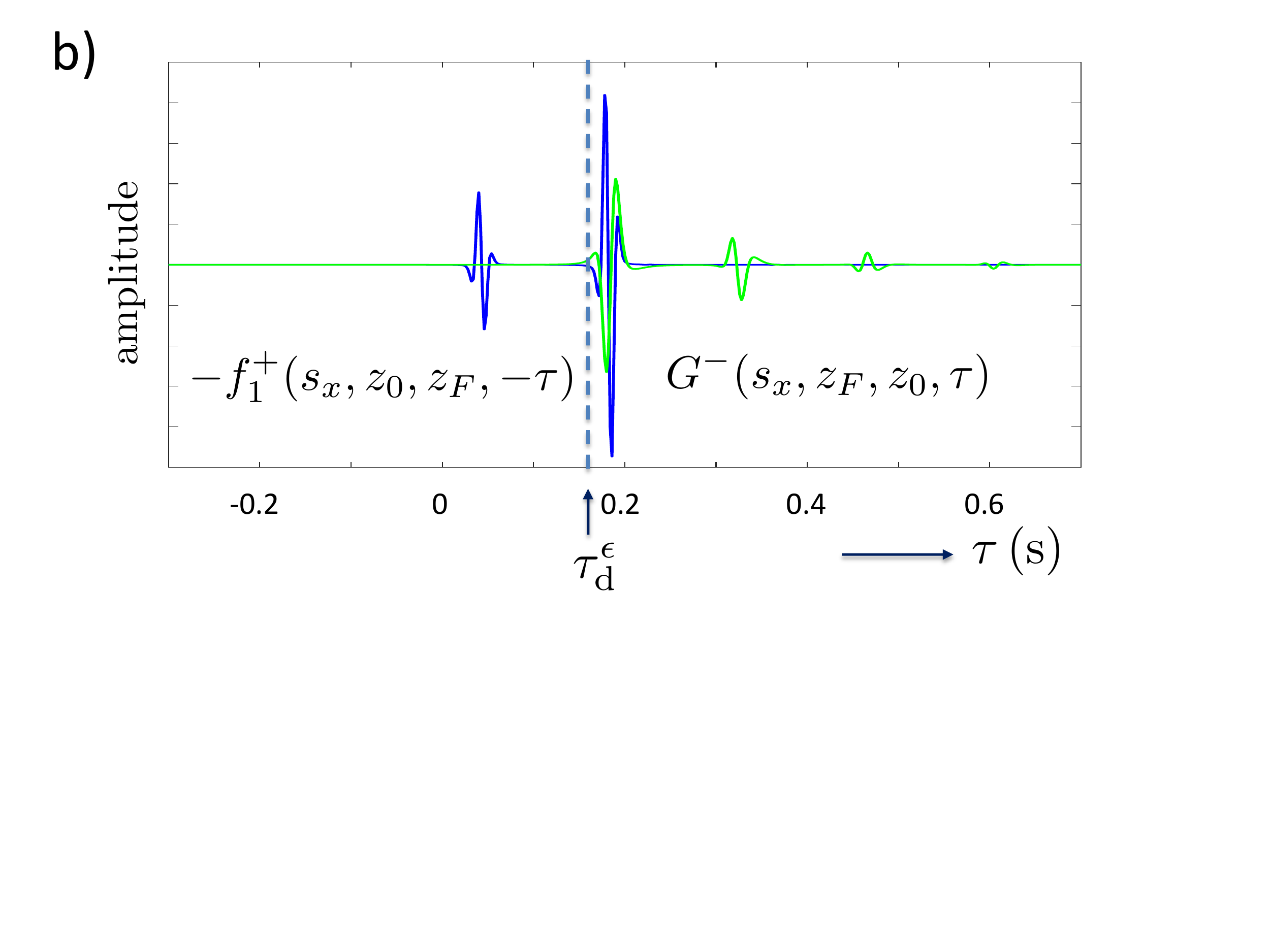}}
\vspace{-2.3cm}
\caption{Functions in the left-hand sides of (a) equation (\ref{eq145}) and (b) equation (\ref{eq148}), for propagating waves at $z_0$ and evanescent waves $z_F$.
In this display the amplitudes of the focusing functions are scaled by a factor 1/8.
}\label{Fig4}
\end{figure}

We discuss some aspects of equations (\ref{eq145}) $-$ (\ref{eq148}).
Consider the medium of Figure \ref{Fig1}, with $z_F=480$ m and $z_3=500$ m. 
Figure \ref{Fig2} shows the functions in the left-hand sides of equations (\ref{eq145}) and (\ref{eq146}),
convolved with a seismic wavelet (central frequency 50 Hz), for $s_x=0$ s/m, hence, for propagating waves at $z_0$ and $z_F$.
The focusing functions are shown in blue and the Green's functions in green. The traveltime of the direct arrival of the downgoing Green's function in Figure \ref{Fig2}(b) is $\tau_{\rm d}$.
The onset of this direct arrival is indicated by $\tau_{\rm d}^\epsilon=\tau_{\rm d}-\epsilon$, where $\epsilon$ is half the duration of the wavelet.
Note that in Figures  \ref{Fig2}(a) and \ref{Fig2}(b), $\tau_{\rm d}^\epsilon$ separates the focusing functions (at $\tau<\tau_{\rm d}^\epsilon$) from the Green's functions (at $\tau>\tau_{\rm d}^\epsilon$), 
except for the coinciding direct arrivals in Figure \ref{Fig2}(b) (equation \ref{eq146}).
This separation is an essential requirement for the standard Marchenko method. 
Next, consider again the medium of Figure \ref{Fig1}, this time with $z_F=420$ m and $z_3=430$ m. 
The third layer between $z_2$ and $z_3$ is now a thin layer.
Figure \ref{Fig4} shows the functions in the left-hand sides of equations (\ref{eq145}) and (\ref{eq148}) for $s_x=1/2800$ s/m, hence, 
for propagating waves at $z_0$ and evanescent waves at $z_F$. 
Note that for this situation there appear to be coinciding arrivals in both equations, hence, the aforementioned requirement for the standard 
Marchenko method is not fulfilled. The mentioned arrivals will remain coincident even when the focal depth $z_F$ is varied within the thin layer, since for evanescent waves the traveltime
does not vary with depth.

To resolve this issue, we derive a relation between $f_1^+$ and $f_1^-$. To this end, we first introduce focusing functions $f_2^+$ and $f_2^-$ \citep{Wapenaar2014GEO}.
The upgoing focusing function $\tilde f_2^-(s_x,z,z_0,\omega)$ is defined such that, when emitted from $z=z_F$ into the truncated medium, it focuses at $z_0$. 
In Table 1 we replace the quantities in state $B$ by $\tilde p_B^\pm(s_x,z_F,\omega)=\tilde f_2^\pm(s_x,z_F,z_0,\omega)$, 
$\tilde p_B^-(s_x,z_0,\omega)=\tilde f_2^-(s_x,z_0,z_0,\omega)=\rho(z_0)/2s_z (s_x,z_0)$ and $\tilde p_B^+(s_x,z_0,\omega)=0$.
State $A$ remains unchanged. From propagation invariant (\ref{prop3})  we obtain (after an inverse Fourier transform)
\begin{eqnarray}
&&  f_1^+(s_x,z_0,z_F,\tau)=  f_2^-(s_x,z_F,z_0,\tau).\label{eq145b}
\end{eqnarray}
From propagation invariant (\ref{prop4}) we obtain for propagating waves at $z_0$ and evanescent waves at $z_F$
\begin{eqnarray}
&&-f_1^-(s_x,z_0,z_F,-\tau)=  f_2^-(s_x,z_F,z_0,\tau).\label{eq148b}
\end{eqnarray}
Combining these two equations yields
\begin{eqnarray}
&&f_1^-(s_x,z_0,z_F,\tau)= -f_1^+(s_x,z_0,z_F,-\tau).\label{eq148c}
\end{eqnarray}
 Using this in either equation (\ref{eq145}) or  (\ref{eq148}) gives
\begin{eqnarray}
&& G^-(s_x,z_F,z_0,\tau) -f_1^+(s_x,z_0,z_F,-\tau)= \int_{-\infty}^\tau R^\cup (s_x,z_0,\tau-\tau') f_1^+(s_x,z_0,z_F,\tau'){\rm d}\tau'.\label{eq145c}
\end{eqnarray}
Hence, for the situation of propagating waves at $z_0$ and evanescent waves at $z_F$, we have reduced the system of equations (\ref{eq145}) and (\ref{eq148}) to 
the single equation (\ref{eq145c}). Since coincident arrivals occur now only in one equation (illustrated by Figure \ref{Fig4}(b)), we have achieved a situation
which can be solved with a modified Marchenko method (to be discussed in the next section). This yields 
 $f_1^+(s_x,z_0,z_F,\tau)$,  $G^-(s_x,z_F,z_0,\tau)$ and (via equation \ref{eq148c}) $ f_1^-(s_x,z_0,z_F,\tau)$.

We still need a representation for $G^+(s_x,z_F,z_0,\tau)$, which we derive as follows. In the original Table 1, we replace the quantities in state $A$ by 
$\tilde p_A^-(s_x,z_F,\omega)=1$, $\tilde p_A^+(s_x,z_F,\omega)=\tilde R^\cap (s_x,z_F,\omega)$, $\tilde p_A^-(s_x,z_0,\omega)=\tilde T^- (s_x,z_0,z_F,\omega)$ and $\tilde p_A^+(s_x,z_0,\omega)=0$.
Here $\tilde R^\cap (s_x,z_F,\omega)$ denotes the reflection response ``from below'' of the truncated medium and $\tilde T^- (s_x,z_0,z_F,\omega)$ its upgoing transmission response.
State $B$ remains unchanged.
From propagation invariant (\ref{prop3}) we obtain, after an inverse Fourier transform, using $s_z (s_x,z_0)\rho(z_F) T^- (s_x,z_0,z_F,\tau)=s_z (s_x,z_F)\rho(z_0) T^+ (s_x,z_F,z_0,\tau)$
\citep{Wapenaar98GEO2},
\begin{eqnarray}
 G^+(s_x,z_F,z_0,\tau)=\frac{\rho(z_0) T^+ (s_x,z_F,z_0,\tau)}{{2s_z (s_x,z_0)}}
+ \int_{-\infty}^\tau R^\cap (s_x,z_F,\tau-\tau') G^-(s_x,z_F,z_0,\tau'){\rm d}\tau'.\label{eq30gg}
\end{eqnarray}
According to equation (\ref{eqTinv}),  $T^+(s_x,z_F,z_0,\tau)$ can be obtained from $f_1^+(s_x,z_0,z_F,\tau)$.
We propose to approximate the unknown $R^\cap (s_x,z_F,\tau)$ by its first reflection, coming from the deepest interface above $z_F$.
Since this is a reflection response for evanescent waves, its amplitude is small and its arrival time is zero, hence it does not require an accurate model.

\section{Marchenko method for evanescent waves}\label{sec4}

We use equation (\ref{eq145c}) as the basis for deriving a modified Marchenko method for the situation of propagating waves at $z_0$ and evanescent waves at $z_F$.
Our first aim is to suppress the Green's function $G^-$ from this equation, so that we are left with an equation for the focusing function $f_1^+$.
 We write this focusing function as
\begin{eqnarray}
f_1^+(s_x,z_0,z_F,\tau)=f_{1,{\rm d}}^+(s_x,z_0,z_F,\tau)+M^+(s_x,z_0,z_F,\tau),
\end{eqnarray}
where $f_{1,{\rm d}}^+$ is the direct arrival and $M^+$ the coda. The time-reversed direct arrival is coincident with the direct arrival of $G^-$, whereas the time-reversed coda is separated in time from
$G^-$, see Figure \ref{Fig4}(b) for an example.
We define a window function $w(\tau)=\theta(\tau_{\rm d}^\epsilon-\tau)$, where $\theta(\tau)$ is the Heaviside step function.
Applying this window to both sides of equation (\ref{eq145c}) gives
\begin{eqnarray}
&&  M^+(s_x,z_0,z_F,-\tau)=- w(\tau)\int_{-\infty}^\tau R^\cup (s_x,z_0,\tau-\tau') f_1^+(s_x,z_0,z_F,\tau'){\rm d}\tau'.\label{eq145h}
\end{eqnarray}
This equation, with $M^+$ replaced by $f_1^+-f_{1,{\rm d}}^+$, can be solved with the following iterative scheme
\begin{eqnarray}
  f_{1,k+1}^+(s_x,z_0,z_F,-\tau)=f_{1,{\rm d}}^+(s_x,z_0,z_F,-\tau)
 -w(\tau)\int_{-\infty}^\tau R^\cup (s_x,z_0,\tau-\tau') f_{1,k}^+(s_x,z_0,z_F,\tau'){\rm d}\tau'.\label{eq145cit}
\end{eqnarray}
The scheme starts with $f_{1,1}^+=f_{1,{\rm d}}^+$, where $f_{1,{\rm d}}^+$ is obtained by inverting the direct arrival of the transmission response of the truncated medium, 
analogous to equation (\ref{eqTinv}).
 Because of the evanescent behaviour of the transmission response, the amplitude of $f_{1,{\rm d}}^+$ grows
rapidly with increasing $z_F$, hence, $f_{1,{\rm d}}^+$ is stable only for a finite depth interval in the layer where waves are evanescent.

Hence, when the reflection response $R^\cup$ and the direct arrival of the focusing function, $f_{1,{\rm d}}^+$, are known,
the iterative scheme of equation (\ref{eq145cit}) yields $f_1^+$. 
Subsequently, equations  (\ref{eq30gg}) and (\ref{eq145c}) yield $G^+(s_x,z_F,z_0,\tau)$ and $G^-(s_x,z_F,z_0,\tau)$. In these retrieved Green's functions, 
$z_F$ indicates the position of a virtual receiver which observes downward and upward decaying 
 evanescent waves, respectively (or, via reciprocity, a virtual source which emits upward and downward decaying evanescent waves).

\begin{figure}
\centerline{\epsfxsize=9 cm \epsfbox{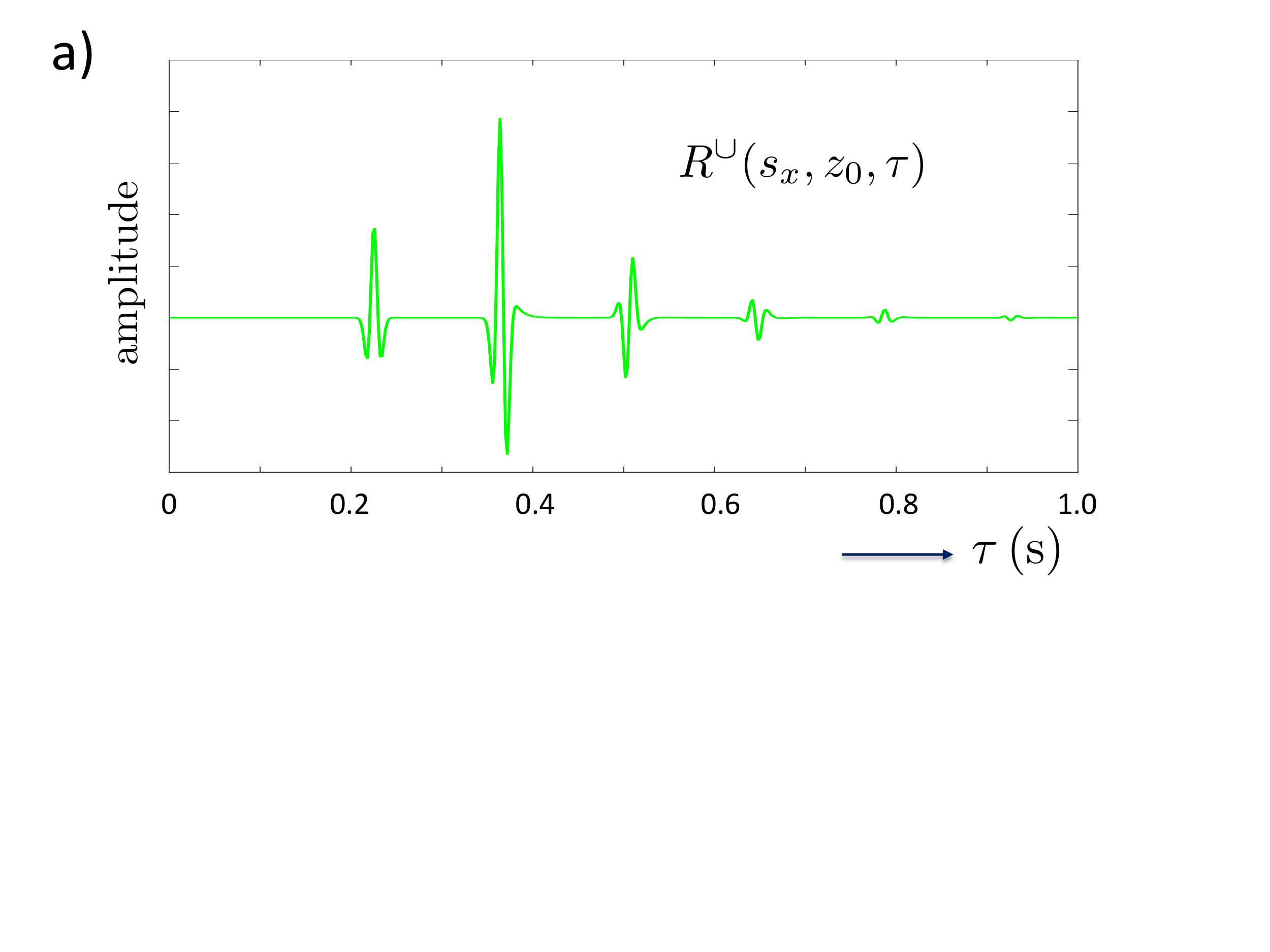}}
\vspace{-2.cm}
\centerline{\epsfxsize=9 cm \epsfbox{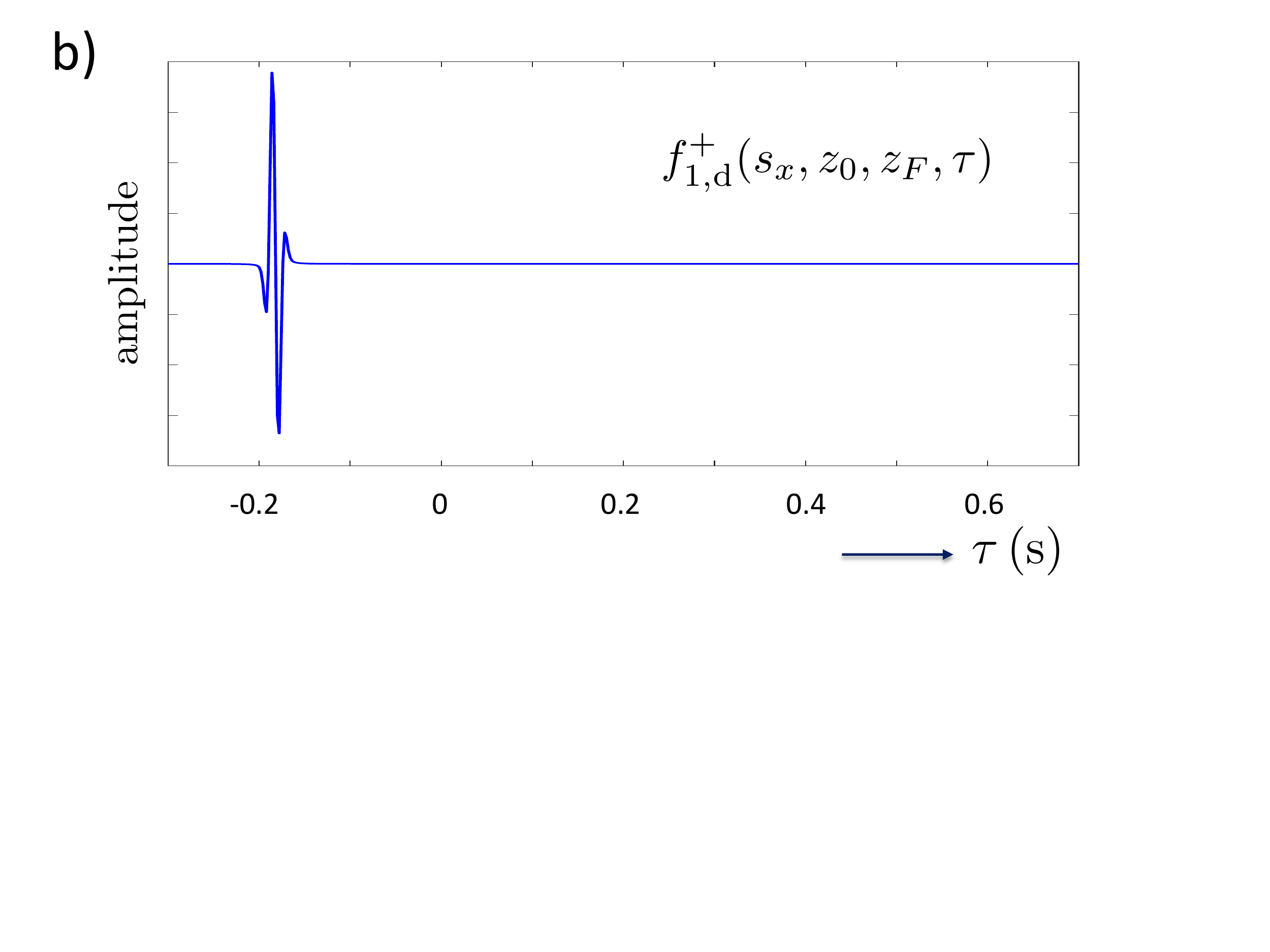}}
\vspace{-2.3cm}
\centerline{\epsfxsize=9 cm \epsfbox{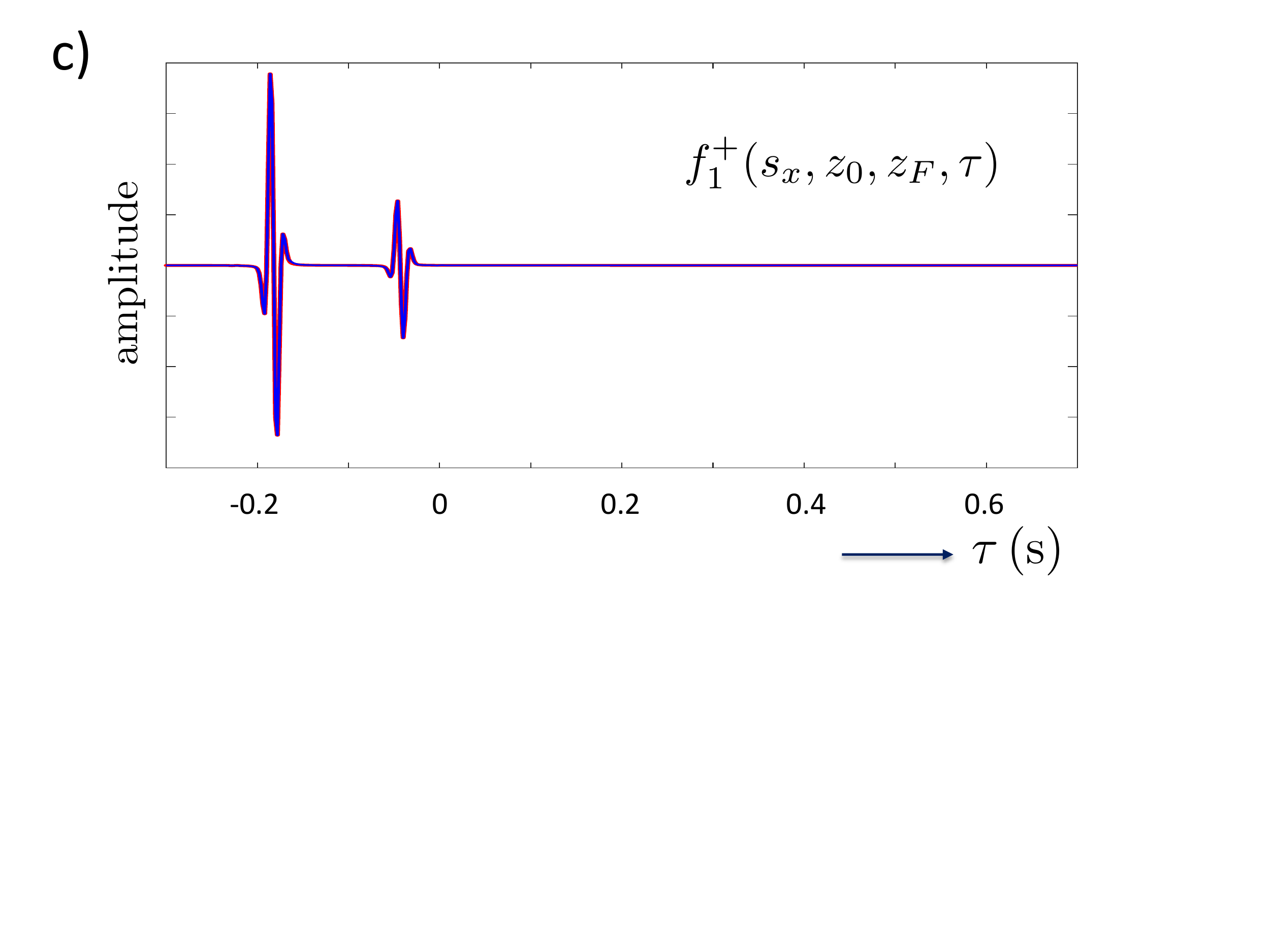}}
\vspace{-2.cm}
\centerline{\epsfxsize=9 cm \epsfbox{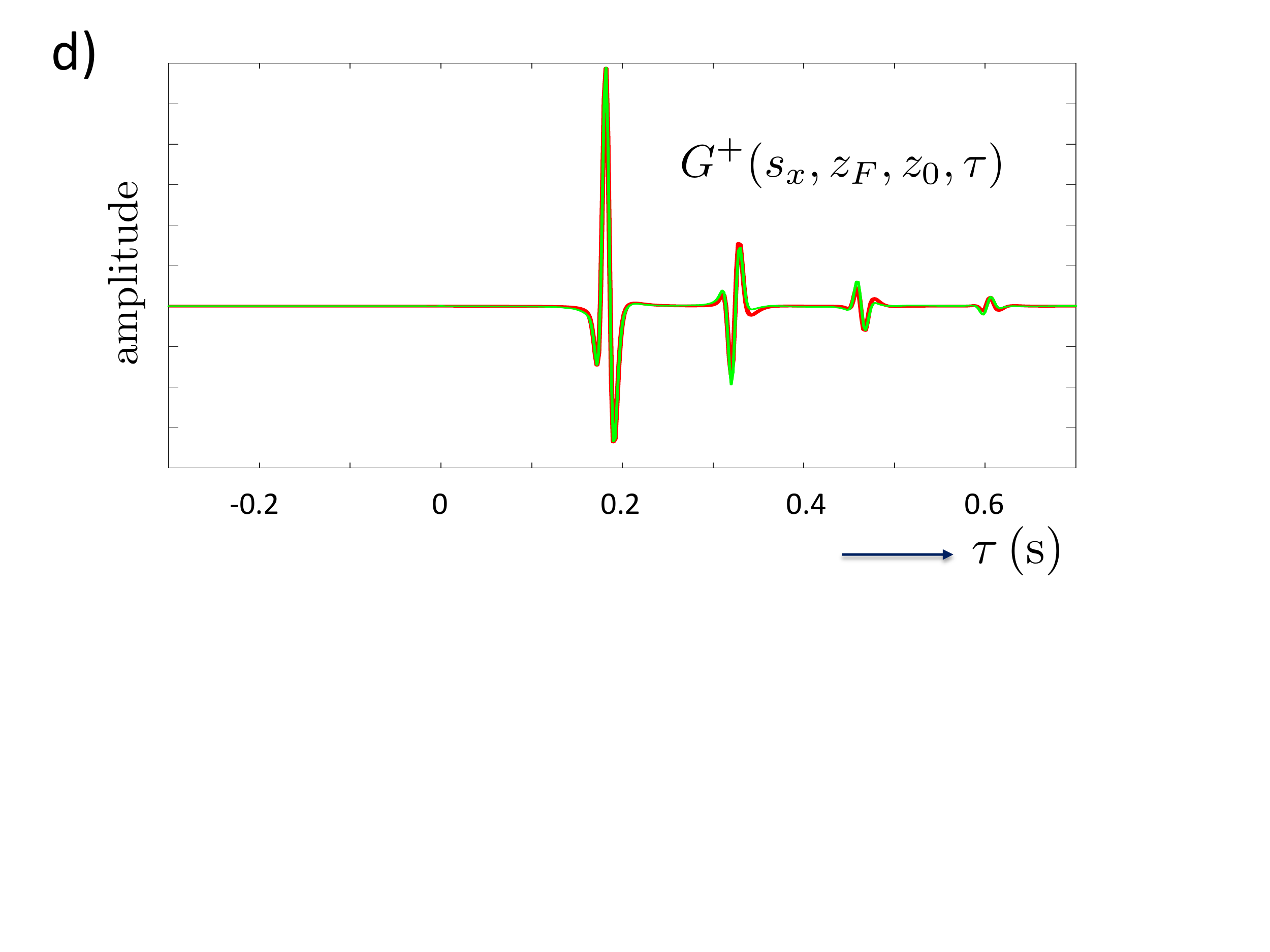}}
\vspace{-2.cm}
\centerline{\epsfxsize=9 cm \epsfbox{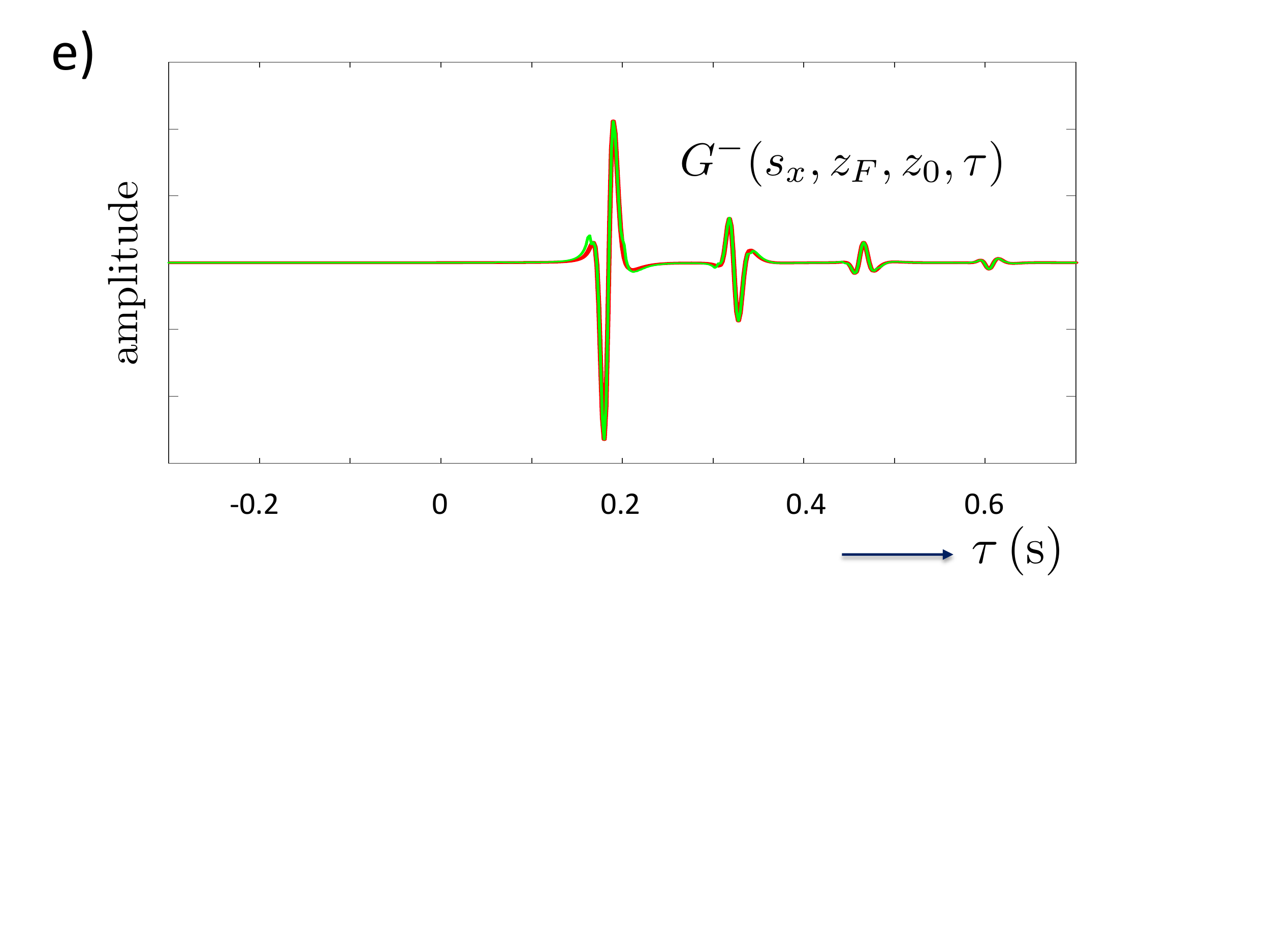}}
\vspace{-2.3cm}
\caption{(a,b) Input data. (c,d,e) Results of the Marchenko method for evanescent waves at $z_F=420$ m.}\label{Fig5}
\end{figure}

\begin{figure}
\centerline{\epsfxsize=9 cm \epsfbox{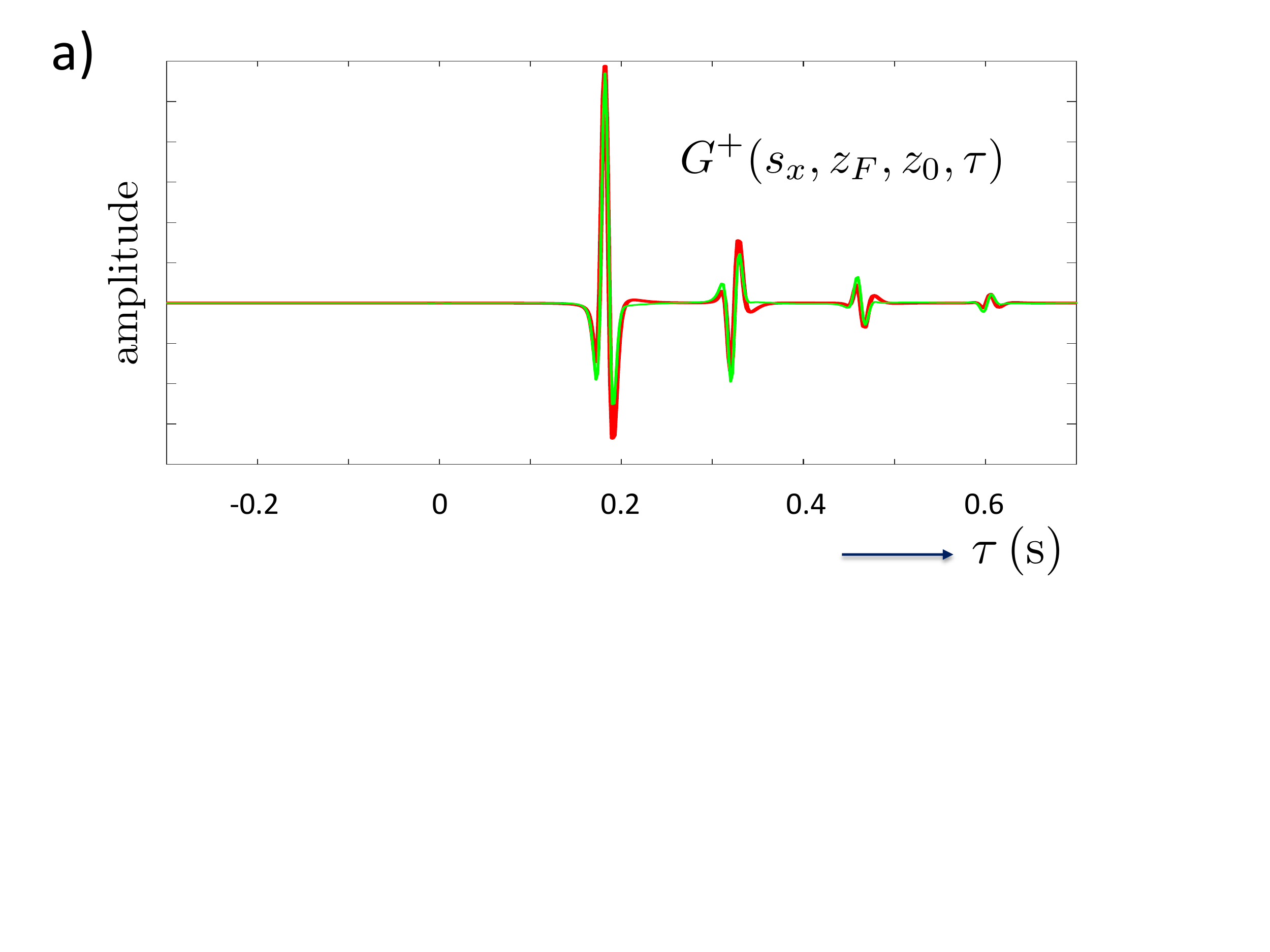}}
\vspace{-2.cm}
\centerline{\epsfxsize=9 cm \epsfbox{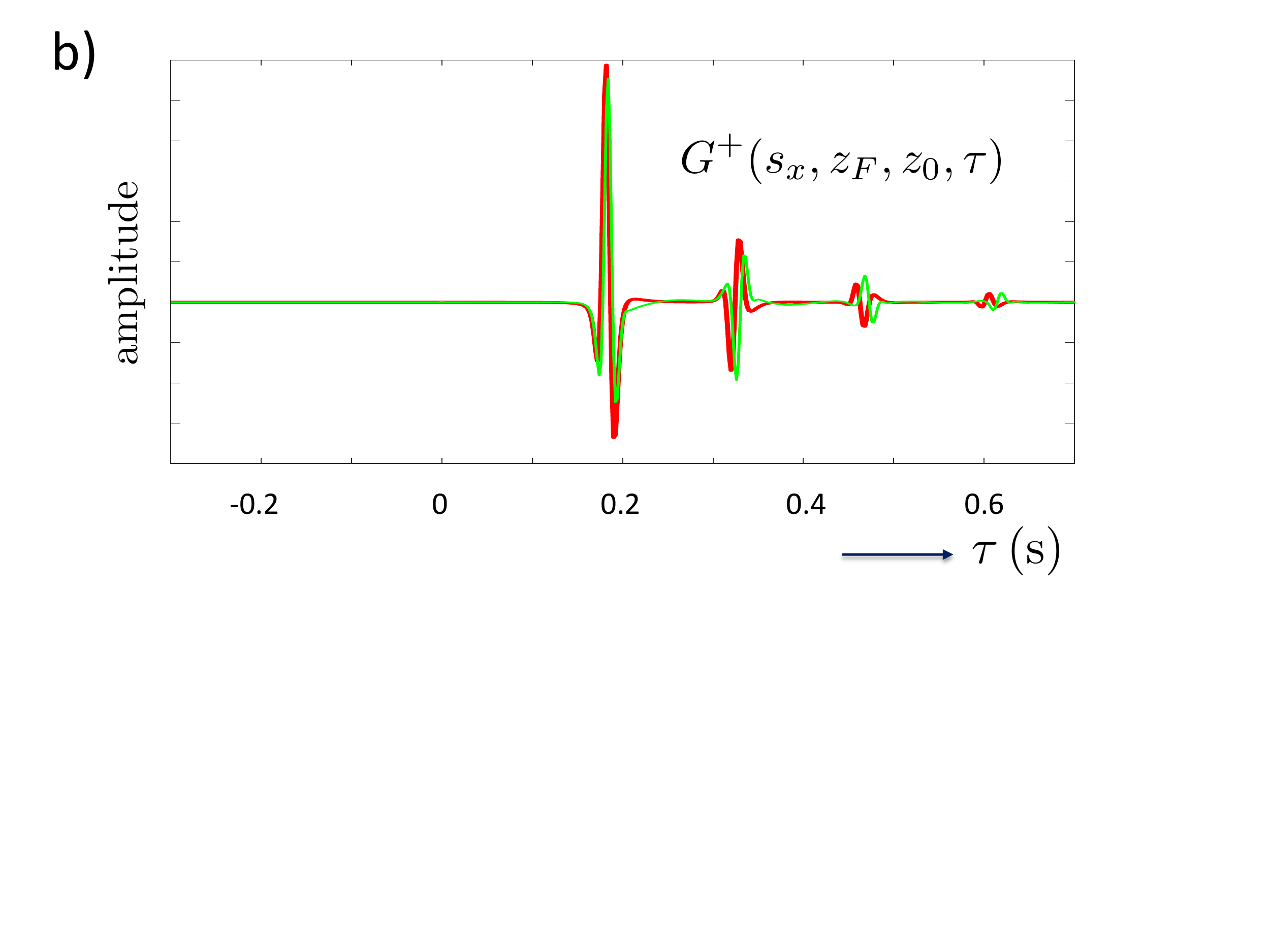}}
\vspace{-2.cm}
\centerline{\epsfxsize=9 cm \epsfbox{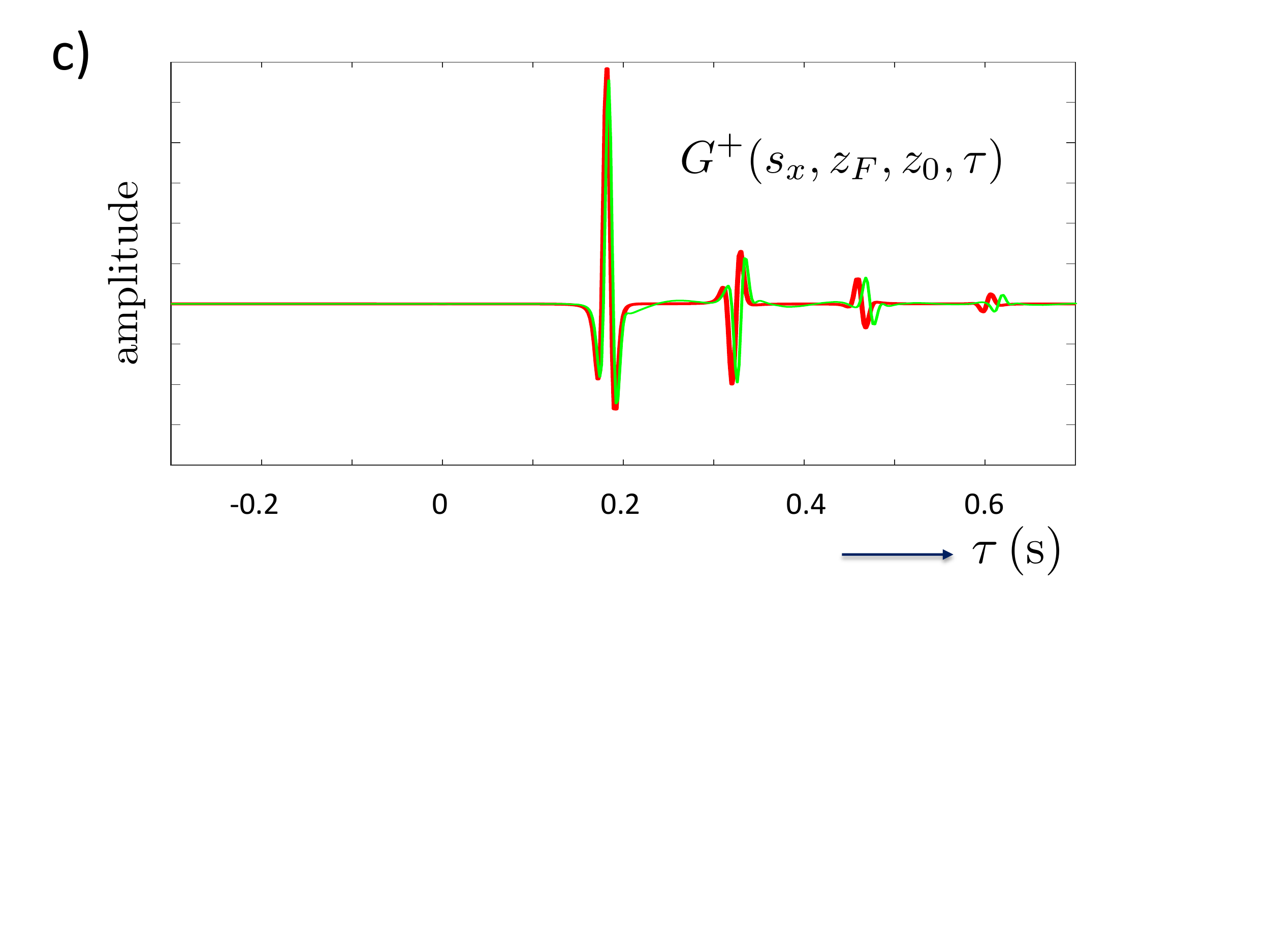}}
\vspace{-2.3cm}
\caption{Results of variations of the Marchenko method  for evanescent waves at $z_F=420$ m (details discussed in the text).}\label{Fig6}
\end{figure}

\begin{figure}
\centerline{\hspace{0.9cm}\epsfxsize=12cm \epsfbox{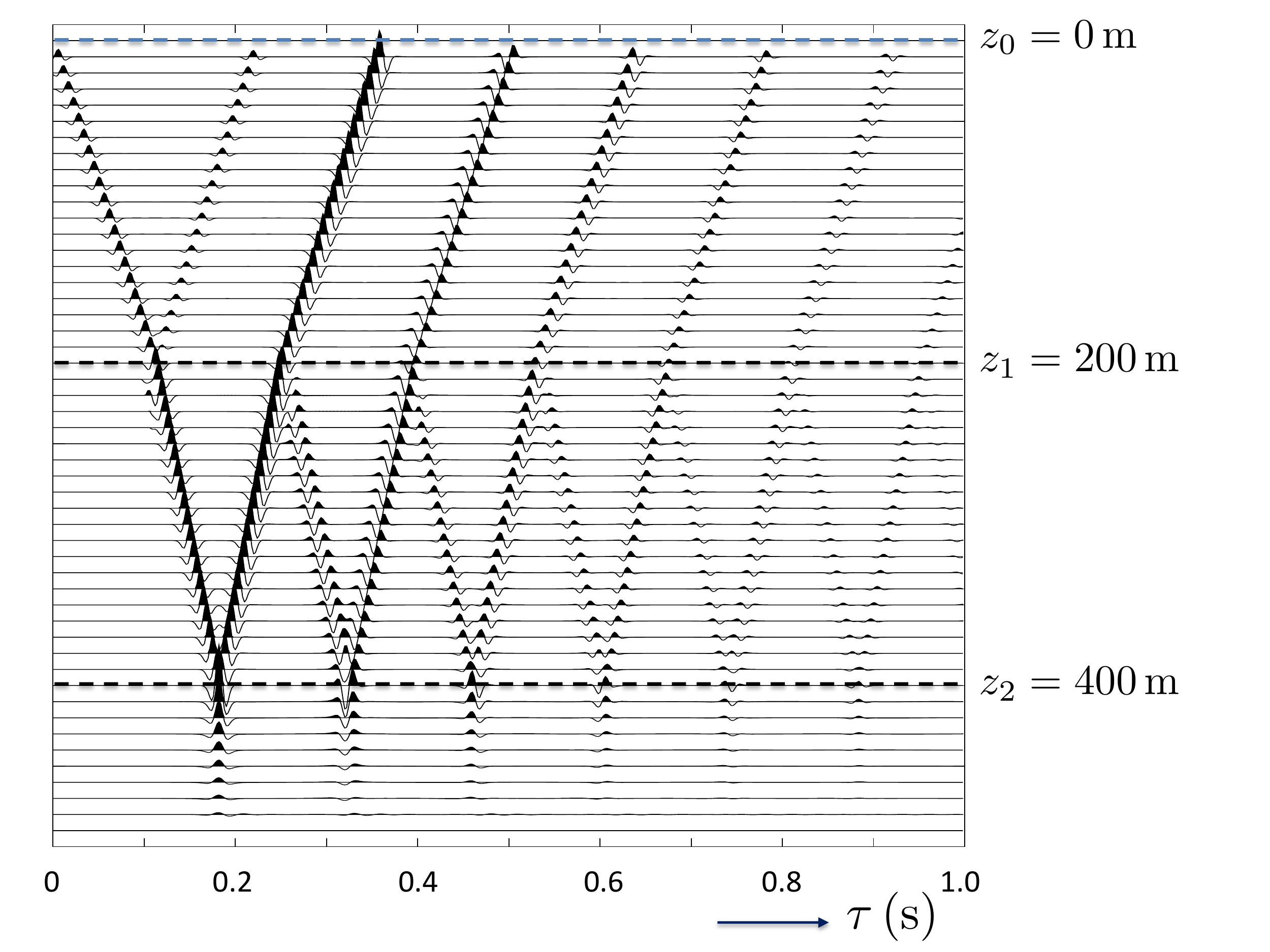}}
\vspace{-.0cm}
\caption{Results of the Marchenko method for all depth levels. To emphasize the multiples, a time-dependent amplitude gain of $\exp\{4\tau\}$ is used in this display. }\label{Fig7}
\end{figure}

We illustrate this for the medium of Figure \ref{Fig1}, again with $z_F=420$ m and $z_3=430$ m.
Figure \ref{Fig5}(a) shows the reflection response $R^\cup (s_x,z_0,\tau)$ for $s_x=1/2800$ s/m.
The direct focusing function  $f_{1,{\rm d}}^+(s_x,z_0,z_F,\tau)$, shown in Figure \ref{Fig5}(b), has been derived from the direct  transmission response, 
modeled for the moment in the exact truncated medium. 
After three iterations, we obtain the results shown in Figures \ref{Fig5}(c,d,e) 
(actually, for this simple medium the method converges already after one iteration and remains stable even after 100 iterations).
The results (shown again in blue and green) overlay the directly modeled exact results (shown in red). Note that the match is excellent (both for the primary and the multiples)
despite the simple approximation used for $R^\cap $, described below equation (\ref{eq30gg}).

Numerical experiments, using erroneous velocities for modeling the direct transmission response, reveal that the method is stable with respect to 
small velocity errors for estimating $f_1^+$, but unstable for estimating $G^-$ (unlike the Marchenko method for propagating waves).  
This means that in practical applications $G^-$ cannot be obtained 
and that the representation for $G^+$ (equation \ref{eq30gg}) should be approximated by the first term.
This obviates the need for estimating $R^\cap (s_x,z_F,\tau)$. Figure \ref{Fig6}(a) shows $G^+$ obtained from the first term in equation
(\ref{eq30gg}). Apart from some amplitude errors, the result is still accurate.
Figure \ref{Fig6}(b) shows again $G^+$, but this time after modeling the direct transmission response in an erroneous truncated medium, 
with velocities $\bar c_1=1450$, $\bar c_2=2050$ and $\bar c_3=3030$ m/s.
We observe similar amplitude errors as in Figure \ref{Fig6}(a) and in addition some traveltime errors caused by the wrong velocities.
Nevertheless, primary and multiples are still clearly discernible and no scattering artefacts related to wrong velocities have come up. Next we 
replace the thin layer by a homogeneous half-space $z>z_2$ (with $c_3=3000$ m/s). Figure \ref{Fig6}(c) shows the retrieved $G^+$ (using the same erroneous truncated medium). 
Since in this situation $G^-$ is absent at $z_F$, the first term in equation (\ref{eq30gg}) suffices to retrieve $G^+$. 
This explains why the amplitudes in Figure  \ref{Fig6}(c) are again very accurate.
Finally we apply the Marchenko method for many focal depths (using the standard method for $z_0<z_F\le z_2$ and the new method for evanescent waves for $z_F>z_2$). 
The result is shown in Figure \ref{Fig7}. Below the interface at $z_2=400$ m we clearly observe the retrieved downward decaying Green's function, including multiple reflections related to
 the overlying medium. For $z_F>480$ m the method becomes unstable and the results have been set to zero.

\section{Concluding remarks}

The analysis in this paper shows that, at least in principle, the evanescent field of the Green's function for a virtual receiver (or via reciprocity a virtual source) inside a layered medium 
can be retrieved from the reflection response at the surface and an estimate of the direct transmission response. 
In theory both the downward and upward decaying components can be retrieved.
However, the retrieval of the upward decaying Green's function is very sensitive to errors in the direct transmission response.
The downward decaying Green's function, including multiple reflections, can be retrieved quite accurately, provided the distance over which the field decays is limited. 
Errors in the direct transmission response cause traveltime errors but do not give rise to scattering artefacts.

The analysis is restricted to a horizontally layered medium and a single horizontal slowness. 
Of course the proposed method can  be applied for a range of horizontal slownesses (for propagating and evanescent waves at one or more depth levels $z_F$). 
Combining this with an inverse transform to the space-time domain,
this enables the monitoring of the space-time evolution of a wave field through a layered medium, similar as in \cite{Brackenhoff2019SE} but including refracted waves.
The generalisation of the proposed method for laterally varying media is subject of current research.

\section*{Acknowledgements}
The constructive comments of Leon Diekmann and an anonymous reviewer are highly appreciated. 
This work has received funding from the European Union's Horizon 2020 research and innovation programme: European Research Council (grant agreement 742703).

\newpage

\bibliographystyle{gji}

\end{spacing}
\end{document}